\documentclass{article}
\usepackage{spconf,amsmath,graphicx}
\usepackage{tabularx}
\usepackage{array}  
\usepackage{comment}
\usepackage{xcolor}
\usepackage{hyperref}
\usepackage{enumitem}

\setlist{nosep, leftmargin=14pt}

\usepackage{mwe} 
\usepackage{graphicx}
\usepackage{subcaption}
\usepackage{caption}
\usepackage{cite}

\title{PRISM: Privacy-preserving Inter-Site MRI Harmonization via Disentangled Representation Learning}

\name{Sarang Galada$^{1,3,*}$ \thanks{$^{*}$ equal contribution} \quad Tanurima Halder$^{1,*}$ \quad Kunal Deo$^{1}$ \quad Ram P. Krish$^{3}$ \quad Kshitij Jadhav$^{2}$} 

\address{$^{1}$ Trust Lab, Indian Institute of Technology, Bombay \\
$^{2}$ Koita Centre for Digital Health, Indian Institute of Technology, Bombay \\
$^{3}$ School of Computing and Data Science, Sai University, Chennai}

\begin{document}

\maketitle

\begin{abstract}

Multi-site MRI studies often suffer from site-specific variations arising from differences in methodology, hardware, and acquisition protocols, thereby compromising accuracy and reliability in clinical AI/ML tasks. We present \textit{\textbf{PRISM} (\textbf{Pr}ivacy-preserving \textbf{I}nter-\textbf{S}ite \textbf{M}RI Harmonization}), a novel Deep Learning framework for harmonizing structural brain MRI across multiple sites while preserving data privacy. PRISM employs a dual-branch autoencoder with contrastive learning and variational inference to disentangle anatomical features from style and site-specific variations, enabling unpaired image translation without traveling subjects or multiple MRI modalities. Our modular design allows harmonization to any target site and seamless integration of new sites without the need for retraining or fine-tuning. Using multi-site structural MRI data, we demonstrate PRISM's effectiveness in downstream tasks such as brain tissue segmentation and validate its harmonization performance through multiple experiments. Our framework addresses key challenges in medical AI/ML, including data privacy, distribution shifts, model generalizability and interpretability. Code is available at 
\url{https://github.com/saranggalada/PRISM}

\end{abstract}
\vspace{-5pt}

\section{Introduction}
\label{sec:intro}
\vspace{-5pt}
Deep Learning (DL) has shown immense potential in the medical field, leveraging the vast amounts of data generated in healthcare settings. Yet, several unique challenges hinder its seamless integration and large-scale adoption. A key obstacle is the fragmentation of medical data across numerous sites \cite{rieke2020future}, necessitating data pooling for robust DL model training and higher statistical power \cite{glocker2019machine}. However, the sensitive nature of medical data often precludes inter-site data sharing, hindering collaborative work \cite{rieke2020future}.

\begin{figure}[htbp]
    \centering   
    \vspace{-5pt}\includegraphics[width=0.4\textwidth]{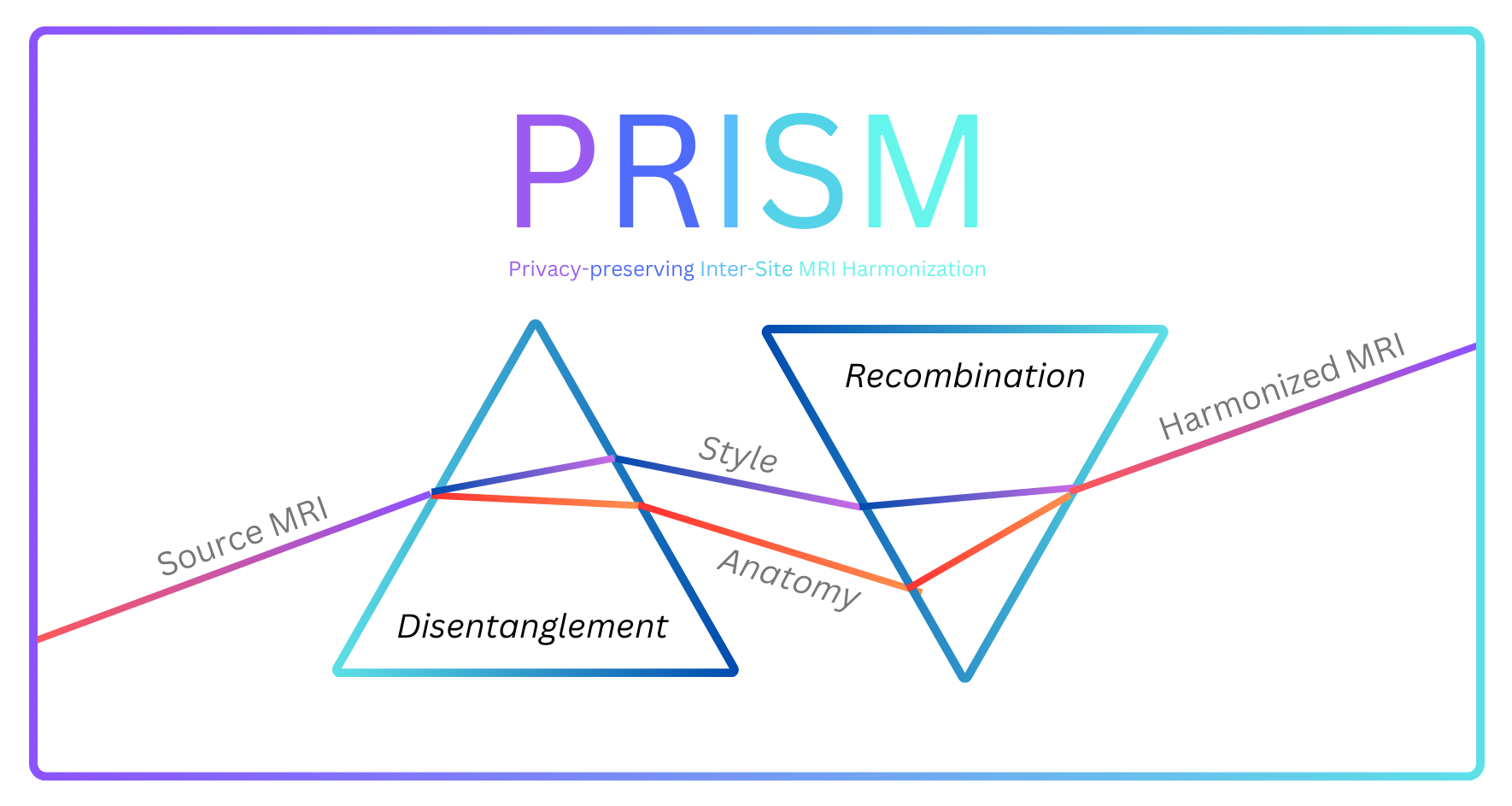}
    \caption{Concept illustration of the PRISM framework}
    \vspace{-15pt}
    \label{fig:prism_logo}
\end{figure}

Federated Learning (FL) \cite{FLmcmahan2017} offers a promising solution, enabling model training across decentralized sites without data sharing \cite{rieke2020future}. However, FL algorithms are particularly susceptible to inter-site heterogeneity and data distribution shifts, which are pervasive in biomedical imaging data \cite{kairouz2021advances}. These ``site-specific effects'' arise from non-biological factors like differences in acquisition protocols and scanner hardware \cite{fortin2018harmonization}. 
Despite efforts to prospectively standardize acquisition parameters, site-specific effects persist, potentially compromising the accuracy, reliability, and interpretability of DL models---factors critical in medical decision-making \cite{glocker2019machine, zuo2023haca3}.

In MRI, site-specific effects manifest as variations in image contrast, resolution, and noise artifacts, leading to unreliable results when applying DL algorithms across different sites or scanners \cite{glocker2019machine}. Moreover, data skew across sites introduces biases, hampering the development of robust, generalizable DL models \cite{kairouz2021advances}. Retrospective harmonization of MRI has been proposed as a solution to mitigate inter-site heterogeneity and non-biological variations \cite{fortin2018harmonization, liu2021style}. Statistical methods, such as intensity histogram matching, empirical Bayes estimation, and z-score normalization, have been widely used, but rely on specific data assumptions and primarily address batch effects rather than complex underlying variations \cite{zuo2023haca3}.

In recent years, deep learning-based harmonization methods have shown superior performance in removing site-specific effects while preserving anatomical information \cite{liu2021style}. These approaches often frame harmonization as an image translation problem, leveraging generative models such as GANs and autoencoders\cite{liu2021style, zuo2023haca3}. While promising, several of these existing methods require paired data or traveling subjects (patients scanned across multiple sites), as well as large amounts of data, which are often unavailable in real-world scenarios \cite{liu2021style}. Additionally, current approaches lack flexibility to harmonize to arbitrary target sites and easily incorporate new sites without retraining, due to their pairwise design.

To address these limitations, we introduce PRISM (\textit{\textbf{Pr}ivacy-preserving \textbf{I}nter-\textbf{S}ite \textbf{M}RI Harmonization}), a modular deep learning framework for harmonizing structural MRI. PRISM approaches harmonization from a content-style disentanglement perspective, leveraging contrastive learning \cite{park2020contrastive} and variational inference \cite{kingma2014auto} to separate anatomical and style information. This disentanglement, coupled with conditional decoding of desired anatomy and style, enables effective multi-site harmonization. Key features of PRISM include:
\vspace{5pt}
\begin{enumerate}
\item Privacy-preserving architecture enabling multi-site collaboration without data sharing.
\item Unpaired image translation capability, eliminating the need for traveling subjects or multiple image modalities.
\item Adaptability to data skew across sites.
\item Flexibility to harmonize to any target site and seamlessly integrate new sites without the need for re-training or fine-tuning, allowing for adaptable harmonization depending on downstream tasks or imaging modalities.
\end{enumerate}
\vspace{-5pt}


\section{Methodology}

\subsection{Data and Preprocessing}
\label{sec:Data_Preprocessing}
In this study, we utilize publicly available structural brain MRI from the \href{https://brain-development.org/ixi-dataset}{\textcolor{blue}{IXI}} and ADNI 
 \cite{mueller2005alzheimer} databases, focusing on T2-weighted scans. Brain extraction was performed using \href{https://open.win.ox.ac.uk/pages/fslcourse/practicals/intro2/index.html}{\textcolor{blue}{FSL BET}} and only the middle axial slice was selected per subject, to ensure model robustness in data-scarce settings. We prepared 3 distinct datasets of axial slices from the sites \textit{Guys} and \textit{HH} (from IXI), and \textit{ADNI}, comprising 319, 185 and 77 images respectively. Preprocessing steps included padding, center-cropping, pixel normalization, background removal and resizing to 256x256 grayscale PNG format. Additionally, for each image, gamma augmentations ($\gamma$=0.5 and $\gamma$=1.5) were included, along with binary masks of the brain region. A $90/10$ train-test split was used across all experiments.
\vspace{-5pt}

\subsection{Model Architecture} \label{sec:Model_Architecture}
\begin{figure}[htbp]
    \centering
    \caption{(a) PRISM’s dual-branch autoencoder architecture. (b) Inter-site harmonization facilitated by encoder-decoder exchange.}
    \begin{subfigure}[b]{0.4\textwidth}
        \centering
        \includegraphics[width=\textwidth]{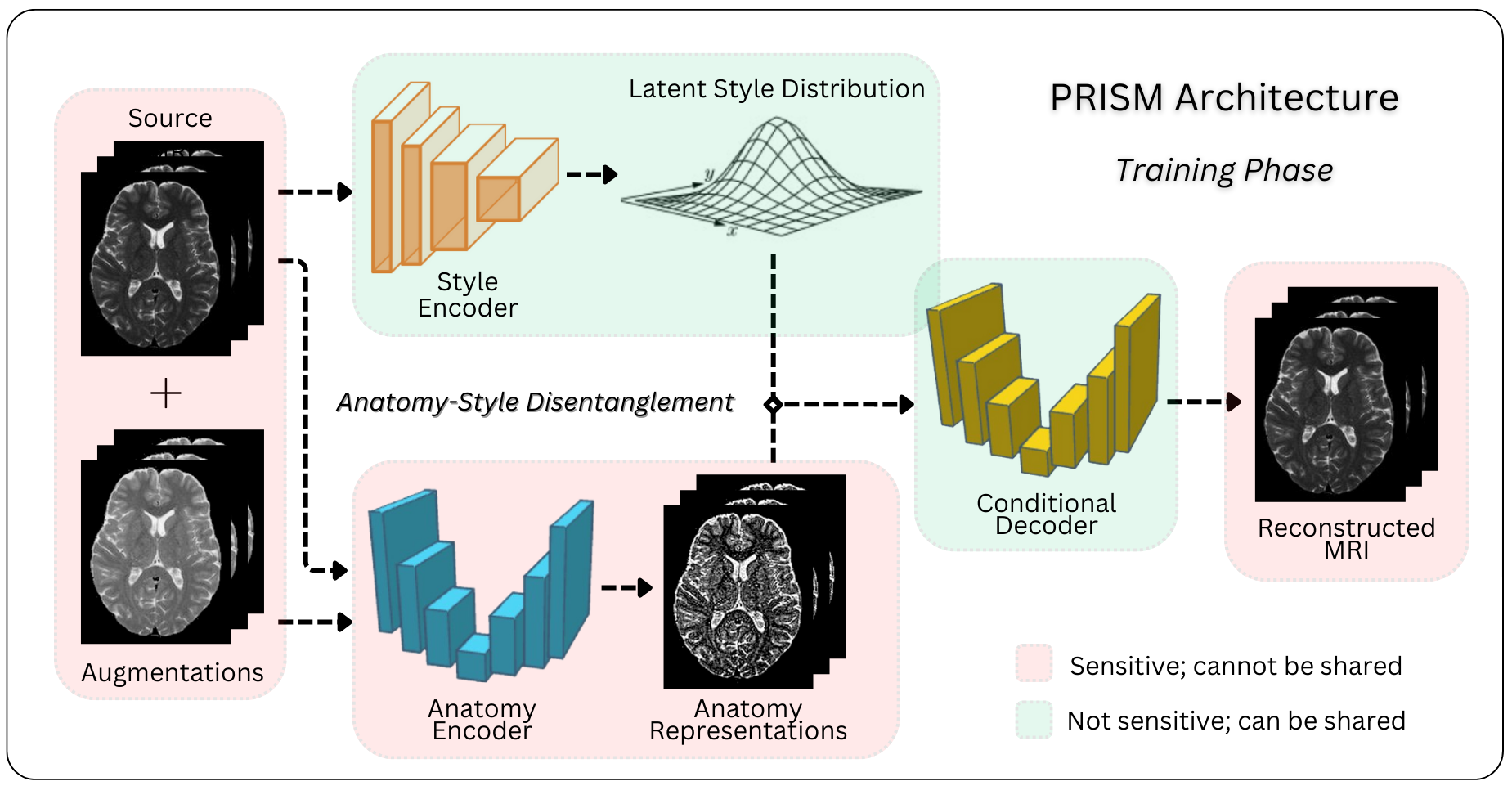}
        \phantomcaption
        \label{fig:prism_train}
    \end{subfigure}
    \begin{subfigure}[b]{0.4\textwidth}
        \centering
        \includegraphics[width=\textwidth]{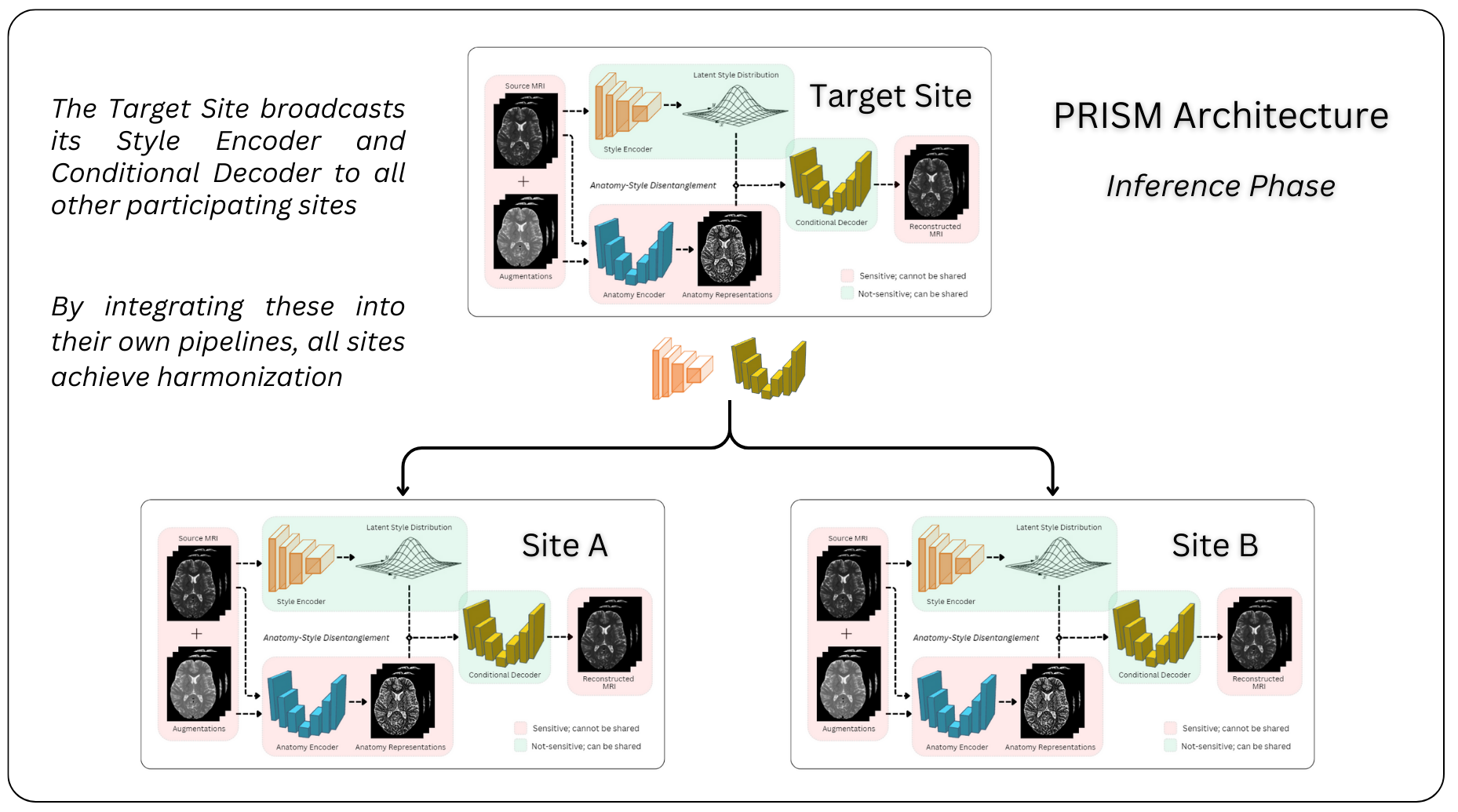}
        \phantomcaption
        \label{fig:prism_inference}
    \end{subfigure}
    \vspace{-7mm}
    \label{fig:prism_combined}
\end{figure}

PRISM’s core architecture is a dual-branch autoencoder designed to disentangle anatomical and style information from 2D MRI slices. It comprises two encoders--for \textit{anatomy} and \textit{style}--and a conditional decoder which recombines the disentangled features to generate images, as depicted in Figure \ref{fig:prism_logo}. Each site locally trains this architecture to reconstruct its MRI scans with high fidelity, learning disentangled representations for \textit{anatomy} (sensitive) and \textit{style} (non-sensitive) (Figure \ref{fig:prism_train}). During inference, the target site broadcasts its style encoder and decoder to all other sites. By using these in place of their own, each site harmonizes its MRI by propagating them through its modified dual-branch autoencoder, without compromising data privacy or anatomical fidelity (Figure \ref{fig:prism_inference}).

Extending on the U-Net \cite{ronneberger2015u} architecture proposed in \cite{zuo2023haca3}, the anatomy encoder takes as input 2D MRI slices along with their gamma augmentations, producing anatomy representations having the same dimensions as the input. Leveraging the \textit{PatchNCE} loss \cite{park2020contrastive}, the encoder maximizes similarity between patches from the anatomical representations produced by source images and their augmentations (query and positive patches) in a projected latent space. This helps capture fine structural details shared between the source images and their augmentations. Parallely, it minimizes similarity between the anatomical patches (queries) and patches from the original source images, as well as unrelated patches from different locations and subjects (negatives). This encourages the encoder to ignore global style-related features.

The style encoder is a convolutional network with a low-dimensional latent space (dim=2), regularized by \textit{KL Divergence} loss to model style and site-specific information as a Gaussian prior, enabling variational inference. Its narrow bottleneck discourages the capture of structural details, focusing instead on global features linked to intensity and contrast.

The conditional decoder leverages a U-Net architecture to reconstruct images, taking as input the anatomy representation with the style representation appended to it as a conditioning channel. Reconstruction is guided by a pixel-wise L1 loss and a perceptual loss. The latter uses a pre-trained VGG16 network \cite{simonyan2014very} to project images onto a latent space, where L1 loss is applied, preserving perceptual quality. Cycle-consistency L1 losses increase alignment between the representations produced by the original and reconstructed images. The model architecture thus formed is a Conditional VAE. Code and implementation details can be found
\href{https://github.com/saranggalada/PRISM}{\textcolor{blue}{here}}.

\section{Experiments and Results}
\label{sec:experiments_and_results}
To evaluate PRISM's performance, we conducted the following experiments/assessments: 1) Reconstruction accuracy on unseen data, to assess model generalizability; 2) Anatomical fidelity of harmonized images, to ensure preservation of critical structural information; 3) Site classification and 4) Visualization of latent style distributions, to quantify and visualize the reduction of site-specific features, respectively; and 5) Brain tissue segmentation as a downstream DL task, to assess the tangible improvements in a real-world application setting.

\subsection{Image Reconstruction and Anatomy Preservation}
Given the limited training data, it's essential that our model reconstructs MRI slices with high fidelity and generalizes well to unseen data. As demonstrated in Table \ref{tab:image_reconstruction_and_anatomy_preservation}, PRISM delivers high-accuracy image reconstructions on the test set. 

To assess the anatomical fidelity of harmonized images, we use SSIM as our primary evaluation metric. Table \ref{tab:image_reconstruction_and_anatomy_preservation} shows that despite the intensity shifts resulting from harmonization, PRISM maintains excellent structural similarity. Notably, our experiments reveal that PRISM is able to accurately reconstruct and harmonize all slices of an MRI volume, highlighting its ability to perform full-volume harmonization.

\begin{table}[htbp]
    \caption{Test data image reconstruction and anatomy preservation in harmonized images. (Values are averaged over the dataset.)}
    \centering
    \footnotesize  
    \renewcommand{\arraystretch}{1.3} 
    \setlength{\tabcolsep}{3.5pt} 

    \begin{tabular}{|c|c|c|c|c|c|c|} \hline 
         \textbf{Site} & \multicolumn{3}{c|}{\textbf{Image-Reconstruction}} & \multicolumn{3}{c|}{\textbf{Anatomy Preservation}} \\ \hline 
         & \textbf{SSIM} & \textbf{PSNR} & \textbf{MSE} & \textbf{SSIM} & \textbf{PSNR} & \textbf{MSE} \\ \hline  
         \textbf{Guys} & 0.9768 & 33.8028 & 0.0004 & 0.9786 & 34.2076 & 0.0004 \\ \hline 
         \textbf{HH} & 0.9775 & 35.4373 & 0.0003 & 0.9598 & 30.9870 & 0.0008 \\ \hline
         \textbf{ADNI} & 0.9524 & 29.7051 & 0.0013 & 0.9266 & 25.5109 & 0.0032 \\ \hline
    \end{tabular}
    
    \label{tab:image_reconstruction_and_anatomy_preservation}
\vspace{-5pt}    
\end{table}

\begin{figure}[htbp]
    \centering
    \caption{PRISM harmonizes tissue intensities and image characteristics, thus reducing inter-site and intra-site variations}
    \includegraphics[width=0.4\textwidth, height=4.7cm]{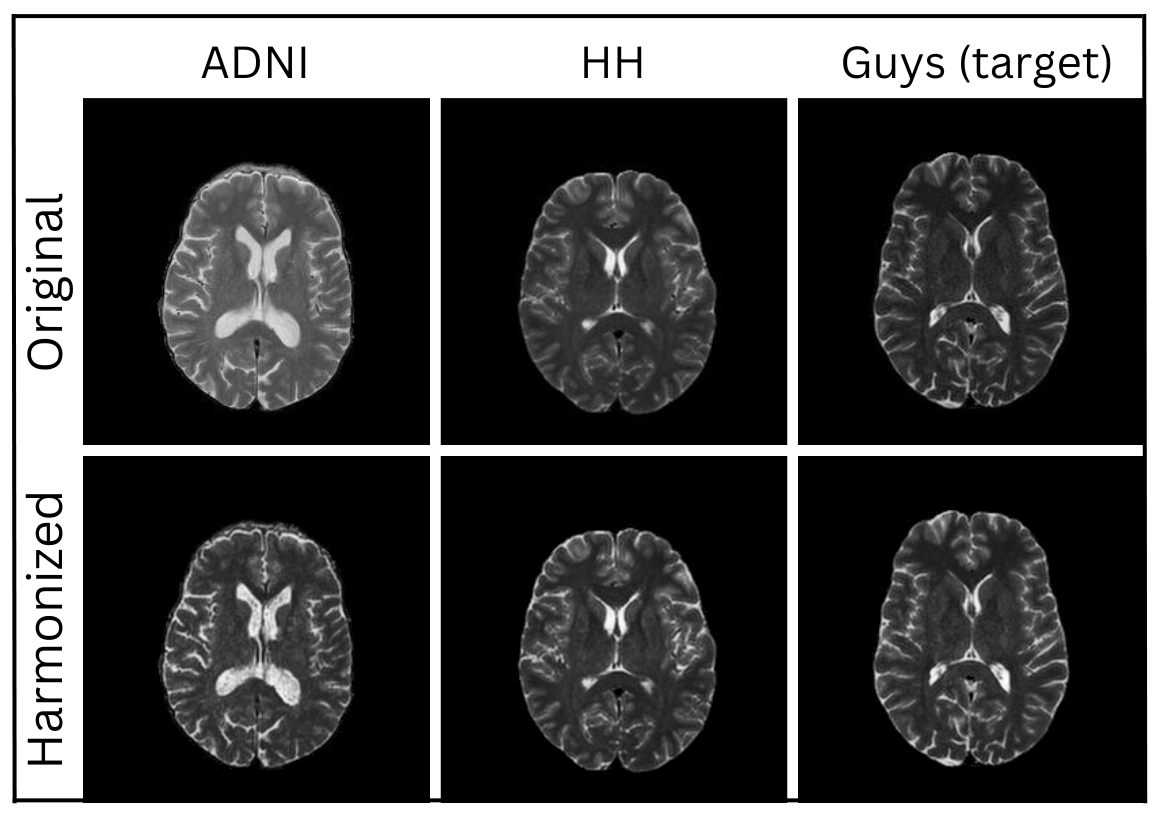}
    \label{fig:harmonized_samples}
    \vspace{-5mm}
\end{figure}

\subsection{Site classification} 
\label{sec:Site_Classfication}
We performed site classification to quantitatively assess harmonization. Two widely used CNN architectures, ResNet50 \cite{ResNet50} and EfficientNet-B2 \cite{Efficient_Net}, were trained on the original (pre-harmonized) datasets to predict the originating site of each image. These models were then evaluated on the harmonized datasets. A significant drop in classification performance, characterized by images being misclassified as the target site, indicates successful reduction of site-specific characteristics. As detailed in \ref{sec:Data_Preprocessing}, we use MRI data from three sites: Guys, HH, and ADNI; choosing Guys as the target site due to its higher SNR. All results are averaged over five independent runs. The same train-test sets are used both before and after harmonization, to prevent data leakage and model bias. 

\begin{table}[h]
    \caption{Site-wise classification (Recall scores averaged over five runs) of original (pre-harmonization) and harmonized datasets.}
    \centering
    \footnotesize
    \renewcommand{\arraystretch}{1.25} 
    \setlength{\tabcolsep}{3pt} 
    
    \begin{tabular}{|c|c|c|c|c|c|c|} \hline 
         \textbf{Model type} & \multicolumn{3}{c|}{\textbf{Pre-Harmonization}} & \multicolumn{3}{c|}{\textbf{Post-Harmonization}} \\ \hline 
         & \textbf{Guys} & \textbf{HH} & \textbf{ADNI} & \textbf{Guys} & \textbf{HH} & \textbf{ADNI} \\ \hline  
         ResNet50 & 0.9935 & 1.0000 & 0.8571 & 1.0000 & 0.3111 & 0.0571 \\ \hline 
         EfficientNetB2 & 1.0000 & 1.0000 & 0.7714 & 1.0000 & 0.6778 & 0.0857\\ \hline
    \end{tabular}
    
    \label{tab:Classifier_Sitewise_Scores}
\end{table}

As shown in Table \ref{tab:Classifier_Sitewise_Scores}, classifiers trained to predict the original site achieve high recall scores on the original test set. However, after harmonization, most images are classified as belonging to the target site (Guys). This trend is reflected in the \textit{weighted F1-scores}, which dropped from \textbf{0.9780} to \textbf{0.5757} for \textbf{ResNet50} and from \textbf{0.9695} to \textbf{0.7383} for \textbf{EfficientNet-B2}. These results signify that our harmonization method substantially reduces site-specific effects.

\begin{figure}[htbp]
    \centering
    \caption{Visualization of each site's latent style distribution.}
    \begin{subfigure}[b]{0.4\textwidth}
        \centering
        \includegraphics[width=\textwidth]{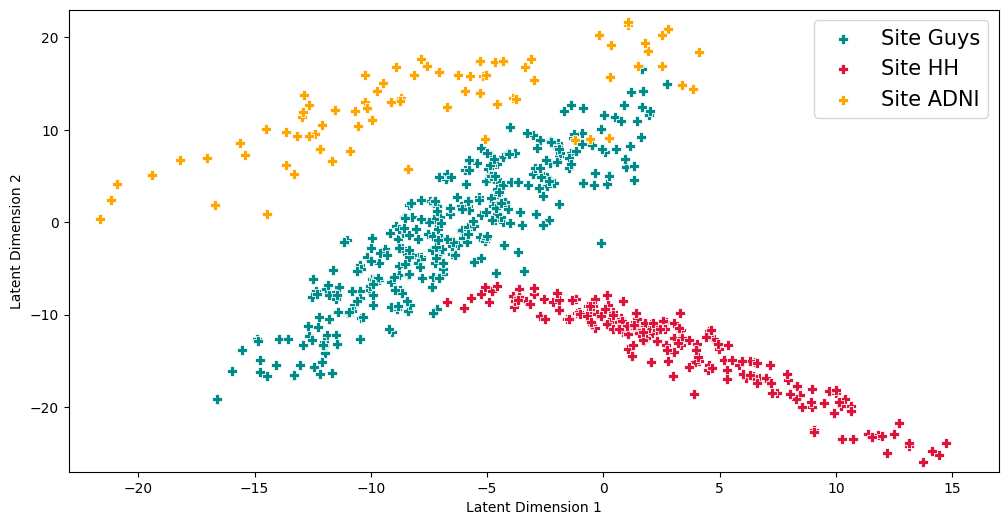}
        \caption{Pre-harmonization}
        \label{fig:style_pre}
    \end{subfigure}
    \hfill

    \begin{subfigure}[b]{0.4\textwidth}
        \centering
        \includegraphics[width=\textwidth]{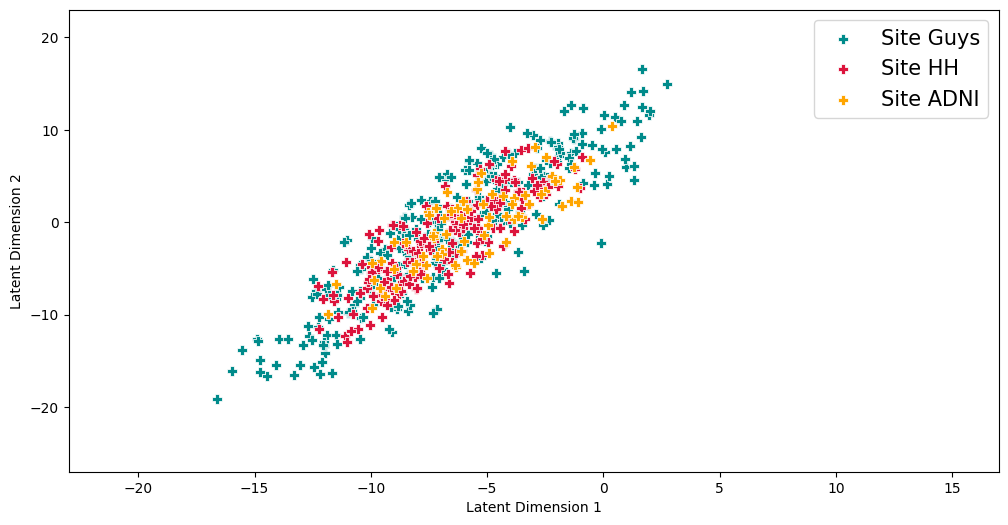}
        \caption{Post-harmonization}
        \label{fig:style_post}
    \end{subfigure}
    \label{fig:style_viz}
    \vspace{-5pt}
\end{figure}

Further validation is provided by visualizing the latent style distributions of each site before and after harmonization, as shown in Figure \ref{fig:style_viz}. PRISM effectively aligns the style distributions of all sites to match that of the target site.

\subsection{Brain tissue segmentation}
The most crucial measure of a harmonization method’s effectiveness is its impact on downstream tasks. To assess this, we chose brain tissue segmentation—segmenting MRI slices into Gray Matter (GM), White Matter (WM), and Cerebrospinal Fluid (CSF)—as our downstream task. 

Ground truth masks for GM, WM and CSF were extracted using the \href{https://www.fmrib.ox.ac.uk/fsl}{\textcolor{blue}{FSL FAST}} pipeline. We employed a U-Net architecture to predict the tissue masks, as it is specifically designed for biomedical image segmentation \cite{ronneberger2015u}. The procedure involved first training the model on images from the target site (Guys), followed by evaluating it on images from other sites (HH and ADNI). Segmentation performance was measured using the Dice Similarity Coefficient (DSC) and Jaccard Index (IoU), with results averaged over five runs.

This procedure was first applied to the original datasets to establish a baseline, and then repeated on the harmonized datasets, to assess the impact of harmonization. As shown in Table \ref{tab:Brain-Tissue-Segmentation-metrics-adni-hh}, harmonization via PRISM significantly improved both DSC and IoU across all tissue types (GM, WM, and CSF), demonstrating its ability to reduce site-specific effects and enhance performance in downstream tasks.

\begin{table}[h]
    \caption{Comparison of brain tissue segmentation metrics (Dice and IoU) on pre-harmonized and post-harmonized datasets of the source sites: (a) ADNI, and (b) HH}
    \centering
    \footnotesize
    \renewcommand{\arraystretch}{1.5} 
    \setlength{\tabcolsep}{4pt}
    
    \begin{tabular}{|c|c|c|c|c|} 
    \hline
    \textbf{Tissue} \textsuperscript{site} & \multicolumn{2}{c|}{\textbf{Pre-Harmonization}} & \multicolumn{2}{c|}{\textbf{Post-Harmonization}} \\ \hline
    & \textbf{Dice coeff.} & \textbf{IoU} & \textbf{Dice coeff.} & \textbf{IoU} \\ \hline
    \textbf{GM} \textsuperscript{a} & 0.8482 & 0.7460 & 0.8614 & 0.7680 \\ \hline
    \textbf{WM} \textsuperscript{a} & 0.8108 &  0.7161 & 0.9087 & 0.8341 \\ \hline
    \textbf{CSF} \textsuperscript{a} & 0.8099 & 0.6737 & 0.8828 & 0.7926 \\ \hline
    \textbf{GM} \textsuperscript{b} & 0.8968 & 0.8154 & 0.9133 & 0.8430 \\ \hline
    \textbf{WM} \textsuperscript{b} & 0.9134 & 0.8425 & 0.9469 & 0.8999 \\ \hline
    \textbf{CSF} \textsuperscript{b} & 0.8314 & 0.7222 & 0.8964 & 0.8158 \\ \hline
    \end{tabular}    
    \label{tab:Brain-Tissue-Segmentation-metrics-adni-hh}
\end{table}

\section{Conclusion and Future Work}
In this work, we introduced PRISM, a modular deep learning framework for harmonizing multi-site brain MRI while preserving data privacy. By careful disentanglement and recombination of anatomy and style, PRISM effectively reduces site-specific variations and improves performance in downstream tasks, such as brain tissue segmentation. Our experiments demonstrate its adaptability to unseen data, ability to preserve anatomical integrity and robustness to data skew. 

Future work will focus on integrating PRISM into Federated Learning downstream tasks (such as Brain-age prediction), and extending it to more number of sites, 3D MRI, and other imaging modalities, to enhance scalability and robustness in decentralized, multi-site medical image analyses.

\section{Acknowledgements}
\label{acknowledgements}
\vspace{-5pt}
The authors thank Trust Lab, IIT Bombay for facilitating and supporting this research.
\vspace{-3pt}

\section{COMPLIANCE WITH ETHICAL STANDARDS}
\vspace{-5pt}
This research study was conducted retrospectively using human subject data made available in open access. Ethical approval was not required as confirmed by the licenses.
\vspace{-2pt}



\bibliographystyle{IEEEbib}
\bibliography{refs}

\end{document}